\documentclass[12pt,a4paper]{article}

\usepackage{essv}
\usepackage{amsmath,graphicx}
\usepackage{listings,acronym}
\usepackage[utf8]{inputenc}
\usepackage{hyperref}

\title{Going Retro: Astonishingly Simple Yet Effective Rule-based Prosody Modelling for Speech Synthesis Simulating Emotion Dimensions}
\author{Felix Burkhardt$^{1}$, Uwe Reichel$^{1}$, Florian Eyben$^{1}$, Björn Schuller$^{1,2,3}$}
\affil{$^1$audEERING GmbH, Germany,\\$^2$Chair EIHW, University of Augsburg, Germany, \\$^3$GLAM, Imperial College London, UK}
\email{fburkhardt@audeering.com}

\begin{document}


\maketitle

\begin{abstract}
We introduce two rule-based models to modify the prosody of speech synthesis in order to modulate the emotion to be expressed. The prosody modulation is based on speech synthesis markup language (SSML) and can be used with any commercial speech synthesizer. The models as well as the optimization result are evaluated against human emotion annotations. Results indicate that with a very simple method both dimensions arousal (.76 UAR) and valence (.43 UAR) can be simulated.
\end{abstract}

\section{Introduction}
\label{sec:intro}

Affect-modulated speech synthesis of a text can be achieved amongst others by
modifying the prosody of the utterance accordingly \cite{scherer1991vocal}. In this work,
emotions will be represented in terms of the dimensional approach of Schlosberg \cite{schlosberg54}, 
who identified the three emotion dimensions
valence, arousal, and dominance. {\em Valence} is referred to as {\em pleasure} in the following.
For this paper, we neglect the dominance dimension for the benefit to focus on the main topic: the control of emotional expression in speech synthesis with a very limited set of prosodic rules.

\subsection{Acoustic correlates of emotions}

For each of these dimensions, several acoustic correlates have been
found. These findings are summarized in \cite{laukka2005dimensional},
\cite{szameitat2011acoustic}, and \cite{mady2021role} (for further
details please see the references therein).

High as opposed to low arousal is characterized by higher speech rate,
higher intensity mean and variability, higher fundamental frequency
($F_0$) mean and variability, higher spectral balance indicating
increased vocal effort, and a higher first formant due to an
increased mouth opening.

Positive as opposed to negative pleasure is amongst others
characterized by higher speech rate and by lower intensity mean and
variability. In addition, pleasure is positively correlated with the
second formant due to more lip spreading caused by smiling
\cite{tartter1980happy}. The relation between pleasure and pitch is
more complicated as found by \cite{banse1996acoustic,
  scherer1991vocal, johnstone2000vocal}: Higher $F_0$ characterizes both
elation joy (positive) and fear (negative), while comfort (positive)
and boredom (negative) are both reflected by lower $F_0$
\cite{scherer1991vocal}.

In general, many results from the literature, for example \cite{msppodcast, Hamada2014, schroeder2004},  indicate that it is difficult to predict and simulate the valence dimension by acoustic cues alone, as opposed to linguistic ones, an assumption that is confirmed also in this investigation.


\subsection{Emotions in speech synthesis}

There are many articles that deal with the simulation of emotional speech and even many that review them, for example \cite{tits2019emotional,Burkhardt2014,schroeder01}; we refer to these for a deeper discussion. 
Historically, first algorithms to simulate emotional  expression were based on prosody rules and categorical emotions. Later, and in line with the new statistical techniques to synthesize speech, data based approaches were used and emotional dimensions as well as speaking styles targeted. Triantafyllopoulos et al.\ review deep learning based approaches in \cite{triantafyllopoulos_2022}.
Marc Schr\"oder was the first on to target emotional dimensions with prosody rules in his dissertation 
\cite{schroeder2004} and his work is one of the foundations of this paper. 
An approach to simulate emotion dimensions with learned features was presented by Hamada \cite{Hamada2014} by mapping acoustic features to the valence-arousal space.
Later, Stanton et al.\ \cite{stanton_2018} 
showed 
how to target the latent space within a Tacotron architecture to generate expressive speaking styles.

\subsection{Emotional simulation with SSML}

Within the scope of the European H2020 EASIER project \cite{morgan:22026:sign-lang:lrec}, we faced the problem to enable a, not-yet emotional but available in many languages, commercial speech synthesizer to simulate emotional expression.
The obvious way to do this, in a way that is agnostic to a specific synthesizer, is to utilize the W3C's Speech Synthesis Markup Language (SSML) \cite{ssml} which is, at least in parts, interpreted by almost all speech synthesis engines that are available. 
This has also been done for categorical emotions by Shaikh et al.\ \cite{shaikh_2009}.

SSML amongst others allows for specifying prosodic modifications of an utterance along the prosodic dimensions pitch, energy, and duration. Thus, speech synthesis can be affect-modulated by mapping emotion dimensions to prosodic parameters based on the findings above, and by passing on these parameters to the Text to speech (TTS) engine via SSML.
We tested this for the commercial Google Speech API\footnote{\url{https://cloud.google.com/text-to-speech}} and the open source MARY TTS engine \cite{mary}, where the support is only partial.

This paper is structured as follows:
In section \ref{sec:models}, we
introduce two rule-based model variants in order to adapt the prosody of an
utterance accordingly. In sections \ref{sec:percept} and \ref{sec:results}, we describe the perceptual evaluation procedure and their results, respectively.
Section \ref{sec:conclusion} concludes the paper with an outlook.

Contributions of this paper are as follows:
\begin{itemize}
    \item We present two approaches to simulate emotional dimensions with SSML, which has to our knowledge not been done before.
    \item We simulate the valence dimension by a very simple pitch manipulation approach.
\end{itemize}

\section{Rule-based affect modulation}
\label{sec:models}

In our study, emotion scores are mapped to speech prosody parameters in two rule-based
algorithms based on the findings introduced in section \ref{sec:intro}. 

\subsection{Method syntact}

As a naive baseline we simply implemented the prosody rules as being positively correlated with pitch and speech rate. To distinguish between arousal and valence, we simply assigned speech rate to arousal and pitch to valence, an approach that worked surprisingly well. 

Of course, we can not be sure if the outcomes are specific to the Google synthesizer that was used to generate the samples. 

\subsection{Method Schroeder}
To try out a more complex approach than Syntact, we implemented a very reduced version of the approach that Marc Schr\"oder described in his dissertation \cite{schroederDiss}. 
To this end, we analyzed Schroeder's MARY TTS \cite{mary} sources \footnote{\url{https://github.com/marytts/marytts/blob/79e4edef3f478dcef0aad3609ba77090e91f0b6d/marytts-client/src/main/resources/marytts/tools/emospeak/emotion-to-mary.xsl}}.
According to the sources, we  extracted the rules displayed in Listing \ref{lst:schroeder_rules}, originally in Java, and implemented them in the Python language. 

\begin{lstlisting}[caption=Prosody rules according to the MARY emotion module]
pitch => 0.3 * arousal + 0.1 *  valence - 0.1 * power
pitch-dynamics => -15 + 0.3 * arousal - 0.3 * power 
range (in Semitones) => 4 + 0.04 * arousal
range-dynamics (min 100) => -40 + 1.2 * arousal + 0.4 * power 
accent-prominence => 0.5 * arousal - 0.5 * valence
preferred-accent-shape => when valence < -20: falling, 
    when valence > 40: alternating, else rising
accent-slope => 1 * arousal - 0.5  * valence
rate => 0.5 * arousal + 0.2  * valence
number-of-pauses => 0.7 * $arousal pause
duration => -0.2 * arousal
vowel-duration => 0.3 * valence + 0.3 * power
nasal-duration => 0.3 * valence + 0.3 * power
liquid-duration => 0.3 * valence + 0.3 * power
plosive-duration => 0.5 * arousal - 0.3 * valence 
fricative-duration => 0.5 * arousal - 0.3 * valence
volume => 50 + 0.33 * arousal
\end{lstlisting}\label{lst:schroeder_rules}

To implement this in SSML, we filtered the list for pitch and  speech rate global values resulting in the two rules shown in Listing \ref{lst:distilled_schroeder_rules}. 
These rules had already been tested in the scope of a project to generate an appropriate robot voice for children with the autistic spectrum \cite{burkhardt:erik_2019}.

The resulting values were then scaled as described in the next Section. Of course, this is only a very small subset of the rules defined by Marc Schr\"oder and this might well be the main reason that this approach did not show to be very successful.

\begin{lstlisting}[caption=Reduced prosody rules according to the MARY emotion module]
pitch => 0.3 * arousal + 0.1 * valence 
    - 0.1 * power
rate => 0.5 * arousal + 0.2 * valence
\end{lstlisting}\label{lst:distilled_schroeder_rules}

\subsection{Mapping from dimensions to rules}
We adapted the emotion to prosody
mapping approach of \cite{burkhardt2022}: 
Values for the emotion dimensions arousal, and pleasure are mapped to the {\em pitch}, {\em rate}, and {\em volume} attributes of
the SSML element $<$prosody$>$. This mapping is carried out in the
following way:

\begin{enumerate}
  \item rescale the scores of emotion dimension $e \in$ \{pleasure, arousal\} to the range $[-1,
    1]$
  \item calculate each of the prosody parameters $y \in$ \{pitch, rate, volume\} by the
      following linear combination
    \begin{eqnarray}
    \label{eq:syntact}
      y & = & \sum_{e\in \{\textrm{pleasure, arousal}\}} w_{e, y} \cdot e
    \end{eqnarray}
  \item rescale $y$ to a range defined by still natural sounding
    minimum and maximum values of the respective prosody dimension.
\end{enumerate}

For both variants, Schroeder, and Syntact, all weights $w_{e, y}$ of emotion dimension $e$ for the calculation of the prosodic dimension $y$ were set manually based on perceptual expert judgments how well the synthesized speech prosodically matches the intended emotions. \cite{burkhardt2022} showed that emotion recognition performance can be increased by adding emotional speech samples synthesized this way to the training data.

For reproducibility, all code is open sourced in the Syntact gitlab repository\footnote{\url{https://github.com/felixbur/syntAct}}.

\section{Perceptual Evaluation}
\label{sec:percept}

We conducted a perception experiment to validate the effectiveness of the approaches. 
We used the Google speech API as a speech synthesizer, with the standard male and female voices. 
As text material, we used two short sentences of the Berlin Emotional Database \cite{emomod}, which are meant to be emotionally undecided:
\begin{itemize}
    \item ''\textit{In sieben Stunden wird es soweit sein.}`` (\textit{it will happen in seven hours.})
    \item ''\textit{Heute Abend könnte ich es ihm sagen.}`` (\textit{i could tell him tonight.})
\end{itemize}
 
The idea is that these sentences are neither too mundane nor already have a linguistic emotional connotation.

\begin{table}[!h]
  \begin{center}
    \begin{tabular}{r|cc}
    Fleiss' $\kappa$ & {\bf Arousal} & {\bf Valence} \\
      \hline
    {\bf Schroeder} & .467 & .067 \\
    {\bf Syntact} & .445 & .121 \\ \hline
    {\bf All} & .461 & .112
    \end{tabular}
    \caption{Fleiss' kappa values for inter rater agreement for arousal and valence levels for methods Schroeder, Syntact, and all values.}
    \label{tab:kappa}
  \end{center}
\end{table}

These four combinations (two sexes times two sentences) were synthesised with both methods and with all combinations (9) of three valence and arousal levels: $.1, .5, .9$, with $.5$ being the neutral level.
resulting in 72 samples ($2\cdot2\cdot2\cdot9$).

The samples were annotated by 10 subjects employed by audEERING GmbH using the I-hear-U-play platform \cite{Hantke15-IIA}.
The labelers were 6 women and 4 men of mean age 34.87 years with 13.79 years standard deviation.
After judging 10 test samples to get acquainted with the task, they answered for each sample the following two questions: 
\begin{itemize}
    \item ''{\it Please rate the arousal level on a scale of  low, mid, and high.}``
    \item ''{\it Please rate the valence level on a scale of negative, neutral, and positive.}``
\end{itemize}
 
To measure the inter-rater agreement we used Fleiss' kappa, the results are depicted in Table \ref{tab:kappa}.  While the raters could agree on the arousal annotations, the values are only slightly agreed on for valence when simulated by the Syntact method but not for the Schroeder method.

\section{Results}
\label{sec:results}

\begin{figure}[t!]
    \includegraphics[width=.46\textwidth]{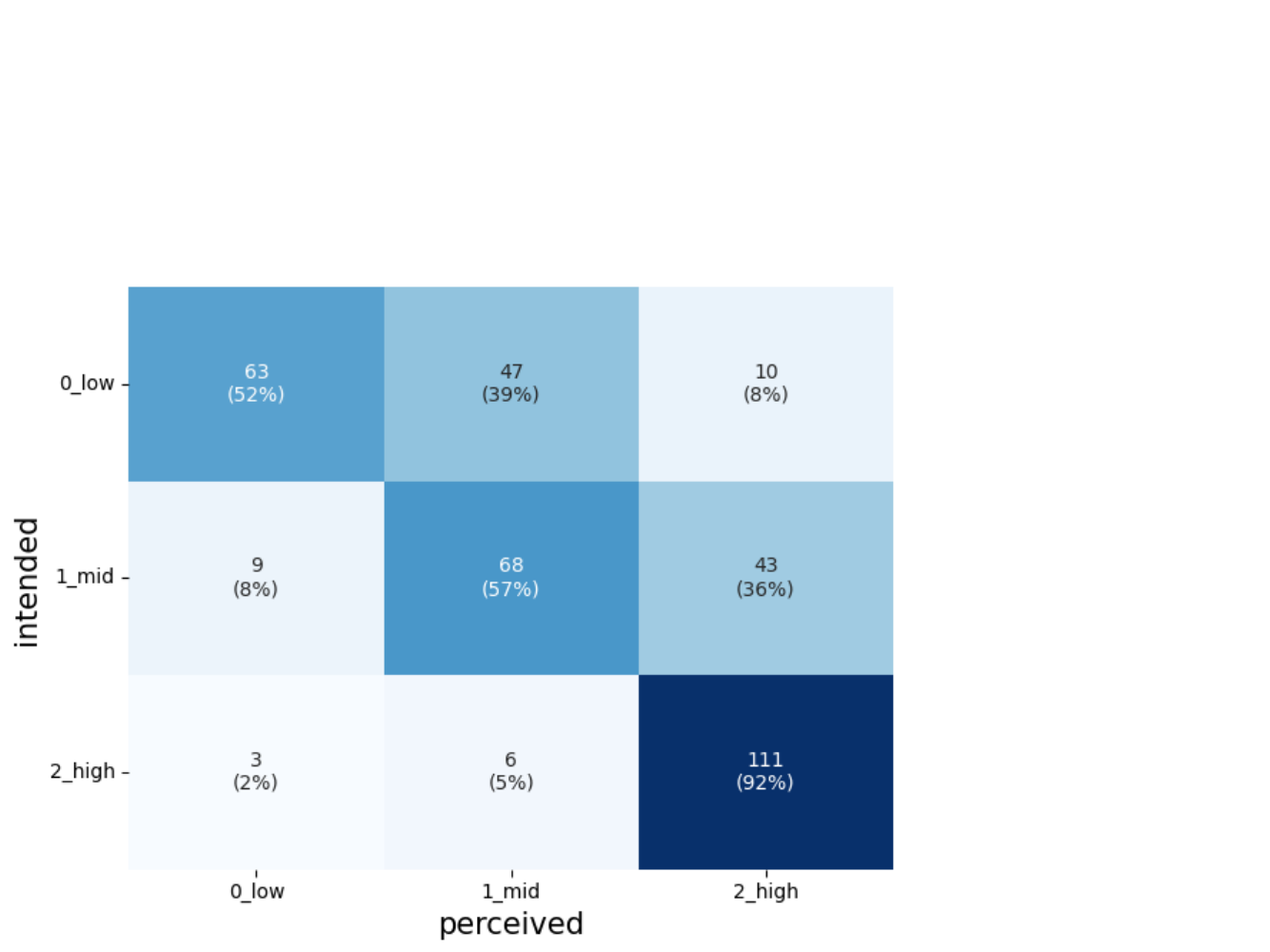}
    \hfill
    \includegraphics[width=.46\textwidth]{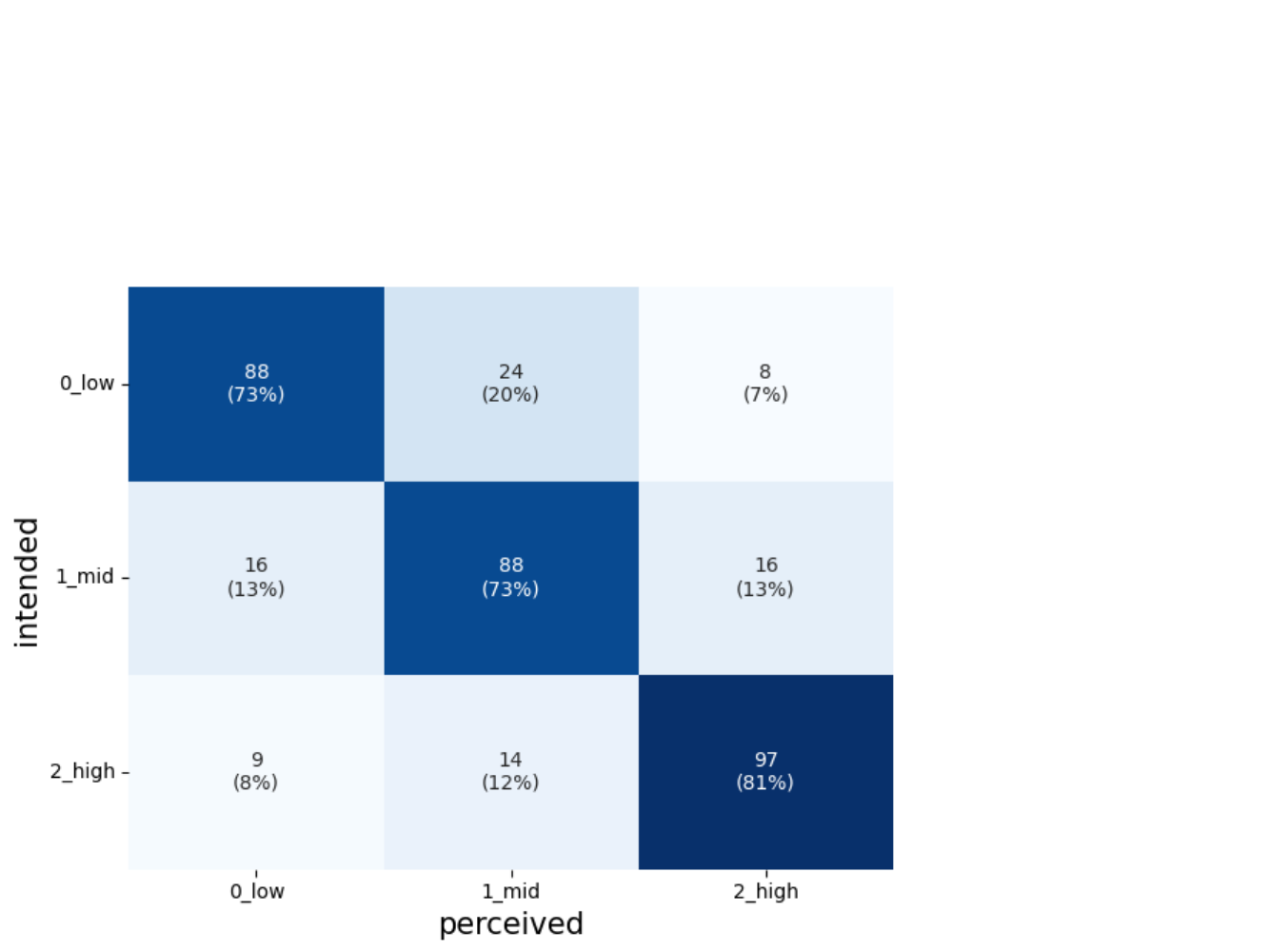}  
    \caption{Confusion matrices between intended and perceived arousal levels. Left: model Schroeder, Right: model Syntact}
  \label{fig:cm_aro}
\end{figure}

\begin{figure}[!ht]
    \includegraphics[width=.46\textwidth]{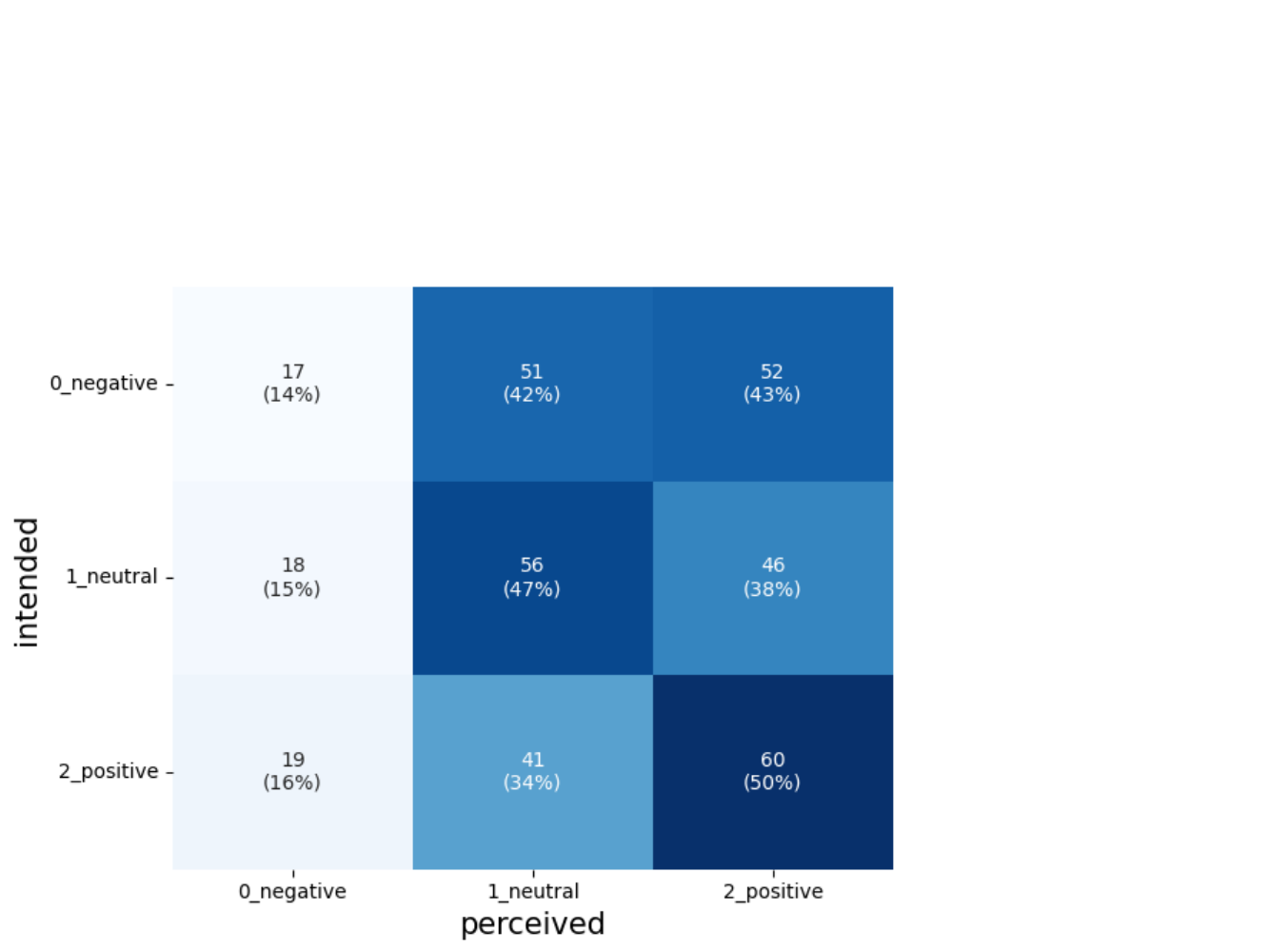}
    \hfill
    \includegraphics[width=.46\textwidth]{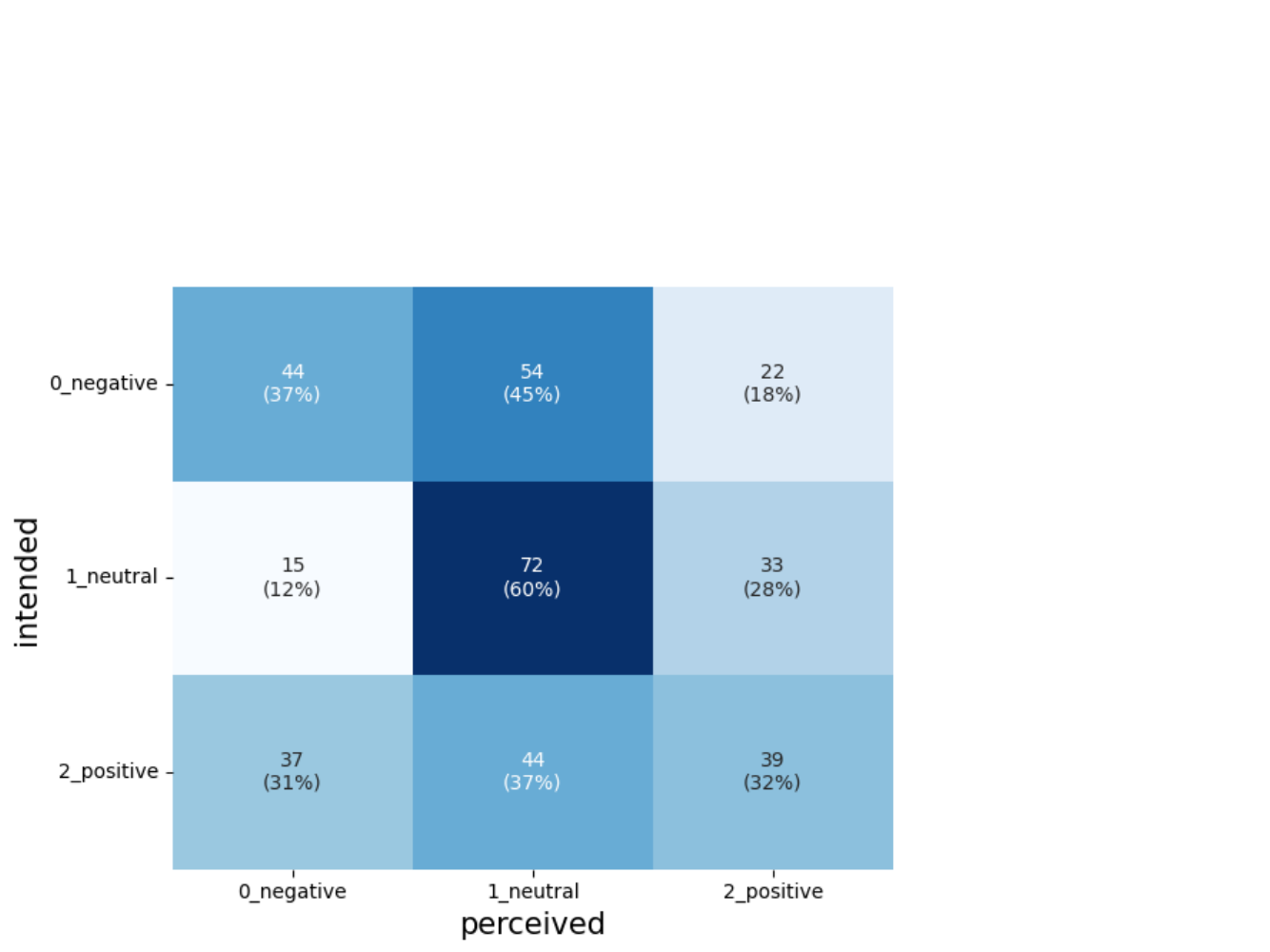}
    \caption{Confusion matrices between intended and perceived valence levels. Left: model Schroeder, Right: model Syntact}
  \label{fig:cm_val}
\end{figure}

The results of the perception experiment are shown in Table \ref{tab:uars}. The confusion matrix for the arousal dimension levels is shown in Figure \ref{fig:cm_aro} and for the valence dimension in Figure \ref{fig:cm_val}. 
The results were computed based on all listeners ratings, without a unified label.

As can be seen, the simulation of arousal was successful with both approaches but to a clearly higher degree with the simpler Syntact method.
For the Schroeder method, low arousal is often confused with the neutral versions and the neutrally meant samples with high arousal.

With respect to valence, we must admit that we were only partly successful with the Syntact method. 
As outlined above, this may largely be also due to the fact that it is mostly convey in linguistic information. This has recently again been shown also for deep representations of acoustics that succeed in better automatic valence recognition from speech due to their inherent encoding of linguistics \cite{Triantafyllopoulos22-PSEa}. 
Nonetheless, we think this is a valuable finding because this method basically hypothesizes that valence correlates positively with pitch which is an interesting approach based on its simplicity.
The samples generated by the Schroeder method were labeled as neutral or high valence by the majority of the listeners, an outcome perhaps that indicates that the ``normal'' expression of the Google voices is rather friendly. 

\begin{table}[!ht]
  \begin{center}
    \begin{tabular}{r|cc}
    UAR & {\bf Arousal} & {\bf Valence} \\
      \hline
    {\bf Schroeder} & .67 & .37 \\
    {\bf Syntact} & .76 & .43
    \end{tabular}
    \caption{UAR values between intended and perceived arousal and valence levels for both methods Schroeder and Syntact.}
    \label{tab:uars}
  \end{center}
\end{table}


As discussed in Section \ref{sec:intro}, it is quite difficult to simulate the valence dimension by acoustic cues alone and accordingly, we are satisfied to have reached even a partial success.

\section{Conclusion and Outlook}
\label{sec:conclusion}
This paper investigated two methods to simulate emotional expression in speech synthesis by controlling prosody with SSML. 
The chosen method following a (strongly) reduced version of Marc Schr\"oder's work did not outperform our rather naive baseline. 

Of course, we cannot be sure if the outcomes are specific to the Google synthesizer that was used to generate the samples. 
Hence, a more general investigation that includes several speech engines will remain future work. 

Also, it is much more promising to learn emotion-to-expression rules from data than to manually determine them based on isolated trials, amongst others because emotional expression is at least to a degree culture specific and the same rules can not be applied in all cultural and social contexts. This has to our knowledge not yet been done for SSML-based approaches and also remains future work.

Thirdly, we restricted the investigation on two dimensions; valence and arousal. Future studies will take at least the dominance dimension into account, which is important to distinguish for example \textit{anger} from \textit{fear}.

It further appears interesting to measure how automatic speech emotion recogniser would recognise such rule- and SSML-based samples. In addition, one could evaluate if they could be used for model augmentation as was first suggested in \cite{Schuller12-SSF}.

On the opposing end -- and likewise closing the circle between analysis and synthesis, one could implement a related rule- and SSML-based recognition of emotion from speech. Presumably, however, this would require some form of speaker normalisation grounded in neutral speech, hence, requiring an enrolment procedure.

\section{Acknowledgements}
\label{sec:ackn}
This research has been partly funded by the European SHIFT (MetamorphoSis of cultural Heritage Into augmented hypermedia assets For enhanced accessibiliTy and inclusion) project (Grant Agreement number: 101060660). 

\bibliographystyle{essv}
\bibliography{essv}

\end{document}